\documentstyle[preprint,aps,prl]{revtex} 
\tightenlines

\begin{document}
\draft

\title{Nanowire Gold Chains: Formation Mechanisms and Conductance} 

\author{
Hannu H\"akkinen, Robert N. Barnett, Andrew G. Scherbakov  and Uzi Landman}
\address{School of Physics, Georgia Institute of Technology, 
Atlanta, GA 30332-0430}
\date{\today}
\maketitle

\begin{abstract}

Structural transformations, electronic spectra
and ballistic transport in pulled   gold nanowires 
are investigated with {\it ab initio} 
simulations, and correlated with recent measurements. 
Strain-induced yield of an initial double-strand
 wire results first  in formation of a bent-chain 
 which transforms upon further elongation
to a linear atomic chain exhibiting dimerized atomic
configurations.
These  structures are stabilized by
directional local bonding with  $spd$-hybridization.
The  conductance  of the initial double-stranded contact is close to 
$2(2e^2/h)\equiv 2g_0$
 and it drops sharply
to $1g_0$ during the transformation to a single chain,
exhibiting  subsequently a $\sim$$1g_0$ plateau extending 
over an elongation  well above  typical Au-Au distances.

\end{abstract}

\narrowtext

Generation of nanometer-scale crystalline wires (NWs) through
elongation of materials' contacts
had been predicted via early molecular dynamics  simulations
\cite{Lan90}, and
their mechanical, structural, and electrical properties have been the subject
of intensive investigations\cite{Ser97}
 owing to increasing  basic and 
technological interest in such NWs. 
Most recently\cite{Ohn98} 
 high-resolution electron microscope  (HRTEM)
images, recorded during 
retraction of a tip from a gold surface, portrayed the development of a 
sequence of crystalline NW structures
made of
 parallel atomic strands oriented along the
NW axis, with the number of strands decreasing one-by-one  
 culminating
in an one-atom wide and four-atom long
linear chain of Au atoms\cite{Ohn98,Ya98}.
 Furthermore, 
the simultaneously measured conductance 
revealed \cite{Ohn98} a stair-case of quantized values (close to
integer multiples of the conductance quantum
$g_0=2e^2/h$) with the disappearance of each atomic strand,
 coinciding at the final pulling stages  with an approximate 
unit ($g_0$) drop in the measured conductance.
These experiments stimulated recent electronic
structure calculations pertaining to single-chains
of gold atoms.\cite{tosatti,Oka99,zigzag,springborg}

Aiming at elucidation of the atomic-scale structural
evolution and transformation mechanisms, and of bonding
 and  transport characteristics at the ultimate
stages of  elongation  of gold NWs, we report
here results from large-scale {\it ab initio} density 
functional  simulations\cite{Note1,Hak99} in conjunction
with conductance calculations\cite{Note2}.
Our findings 
reveal that the  formation mechanism  of a
four-atom long  single chain NW,
generated through straining of a double-strand (ds) wire (see bottom left
configuration in Fig. 1a), 
involves stress accumulation followed by eventual yield
of one of the atomic strands, with the consequent
atomic rearrangement resulting first in a
(lower-energy)  bent-chain (bc) configuration
(top left  configuration in Fig. 1a).
The yield event is accompanied by a significant decrease in the pulling
force 
 and  a sharp drop in the conductance to $1g_0$
 (see top  in Fig. 1a). 
Further separation of the two electrodes results in
a transformation to a linear-chain
(lc) structure (see right configuration in  Fig. 1a),
with continued elongation leading to  formation of
dimerized configurations.         
 Initial dimerization develops 
 between the inner atoms
of the wire (i.e. inner-dimerization, id),
transforming at a later stage
 to one where each of the inner
 atoms pairs with an end-atom
of the chain (i.e. end-atom dimerization, ed).
During the ds$\rightarrow$bc$\rightarrow$lc evolution
the conductance exhibits a $\sim$$1g_0$ extended
(4 \AA\ -- 4.5 \AA\ long) plateau 
(see Fig. 1a). 
The optimal structures determined here
are stabilized by directional  local bonding involving
$s,p$ and $d$ hybridization near the Fermi energy, and they
are found to be energetically favorable to certain
suggested alternative structural models.\cite{zigzag}

We start  from a relaxed
ds wire configuration with the
distance between the two outermost
layers of the opposing Au electrodes held at
$L=11.5$ \AA; each of the electrodes is composed of 29 atoms
arranged in 3 layers with face-centered-cubic (110) stacking
in the $z$-direction (along the
wire)  and exposing $\lbrace 100\rbrace$ and $\lbrace 111\rbrace$ 
side facets.
This crystallographic orientation, and the subsequent elongation
process and conductance characteristics discussed below, correspond
to experiments discussed in Ref. 3 and described in Fig. 3 of that
paper, where HRTEM and conductance measurements pertaining to
gold nanowires created via controlled elongation of the contact
between a tip and a (110) facet of a gold island are displayed.
Note that this experiment is distinctly different from the one
corresponding to Fig. 4 of Ref. 3, which pertains to a wire
formed through (spontaneous) thinning of a "nanobridge" connecting
across a hole formed in a gold film by intense electron
bombardment (in particular, unusually large interatomic
distances of up to 3.5 \AA\  -- 4 \AA\ were reported in Ref. 3
for the latter case, and {\it not} for the tip-pulling
experiments which are the focus of our study).

In this initial configuration the 2-atom strands are parallel
to each other, with almost equal inter- and intra-strand separations,
$d_{intra}(1',2')=d_{intra}(1,2)=2.66$ \AA\ and
$d_{inter}(1',1)=
d_{inter}(2',2)=2.67$ \AA\ (see atom
numbering in Fig. 1a),
and the distance  between the topmost (interfacial) facets of the
opposing electrodes is $d_{el-el}=6.02$ \AA.
Note that all the distances involving atoms bridging
the  electrodes 
 are smaller than the nn interatomic distance in bulk
gold (2.885 \AA), but larger than the bond-length of the free
Au$_2$ dimer (2.48 \AA). 
The calculated conductance \cite{Note2} of 
this  configuration is  
$1.79g_0$\cite{longds}.

Elongation of the wire is simulated through increasing 
the separation between the outermost layers of the 
opposing electrodes (where the atoms are
held at their fcc lattice positions with a lattice constant
of bulk gold, 4.08 \AA),
followed by  full relaxation of the system after each
elongation step. The total elongation simulated here is $\Delta L=6.5$
\AA\ (using elongation increments of
0.25 \AA $\leq \delta L\leq$ 1 \AA, see
Fig. 1a), and we found that   
 structural changes in response
to increments in $L$ involve mainly atoms 
 forming the connective wire  bridging 
the electrodes.

{\it ds$\rightarrow$bc transformation}\ ($L\leq 13.3$ \AA).\  \ \ 
Initial elongation results  in
higher-energy strained configurations with the pulling forces 
rising to $\sim 4$ nN (corresponding to
the elongation increment from
$L=12.3$ \AA\ to 12.6 \AA, see
Fig. 1a), and it is accompanied by a gradual (small)
decrease of the conductance 
to $G=1.68g_0$ in the highly
strained ds configuration shown at the bottom left
in Fig. 1a (corresponding to $L=12.6$ \AA\ with
$d_{el-el}=6.95$ \AA). 

For all the ds configurations the ballistic
electron transport involves mainly two conductance eigenchannels
(CCs), with the total conductance $G=g_0\sum_n\vert\tau_n\vert^2$,
where $0\leq\vert\tau_n\vert^2\leq 1$ is the
 transmission probability 
of the $n$th eigenchannel; for the initial
ds configuration $\vert\tau_1\vert^2=0.99$,
$\vert\tau_2\vert^2=0.76$, and for the highly-strained
one $\vert\tau_1\vert^2=0.97$, $\vert\tau_2\vert^2=0.70$. 
The CCs are determined by quantization of the electron
motion transverse to the propagation direction ($z$),
due to the lateral confinement 
 by the effective wire potential\cite{Hak99,Note2}.
Each of the transmitted CCs is found to be delocalized 
over the two atomic strands with the first CC 
being nodeless, and the second having a nodal
 plane normal to the  plane defined by the
two strands and bisecting the interstrand Au-Au bonds
(see Fig. 2a); the delocalized nature of the
CCs contrasts  the conjecture\cite{Ohn98} that
in such multi-strand NWs each of the
channels is associated with (and is confined about)
an individual atomic strand implying two independent CCs
connected in parallel to the  electrodes.

The ds$\rightarrow$bc
 elongation stage culminates in breaking of one of the
strands, accompanied by displacements of each of the
end atoms of the broken strand (1' and 2' in Fig. 1a) to the
center of the underlying (110) rectangular facet of the corresponding 
electrode, yielding a lower-energy  bc configuration,
  with a concomitant sharp drop in the conductance to  1 $g_0$
 (see Fig. 1a) involving a single CC.
In the relaxed  bc configuration (top left structure in
Fig. 1a) the end-to-end wire length $l_{ee}\equiv d(1',2')=5.12$ \AA,
the distance between an interior atom to an nn  end-atom  of the
wire
$d_{ie}=d(1,1')=d(2,2')=2.62$ \AA, the bond length between the
internal atoms of the wire $d_{ii}=d(1,2)=2.56$ \AA, and the bond
angle $\angle(1',1,2)=119^\circ$. 

{\it bc$\rightarrow$lc transformation} 
(13.3 \AA\ $\leq L\leq$ 16.3 \AA).\ \ \  
During subsequent elongation, requiring initially a small
pulling force ($\sim$0.8 nN, corresponding to
the interval between $L=13.3$ \AA\ and $L=13.8$ \AA), 
 the bc wire straightens gradually resulting
eventually (see right configuration in Fig. 1a)
in a linear chain at $L=16.3$ \AA\ \cite{nootti1}.
The bc$\rightarrow$lc transformation spans an elongation
range $\Delta L(bc\rightarrow lc)= 3.0$ \AA, during which
the conductance decreases only by about 10\% 
(from 1 $g_0$ for the initial bc configuration to 0.87 $g_0$ for
the lc one at $L=16.3$  \AA).

In light of a recent report pertaining to alternative structures
for single-chain gold nanowires\cite{zigzag} 
 we have explored such configurational
isomers during the bc stage, with two of them
shown in Fig. 1b; a planar 4-atom rhombohedral (rh) arrangement
at $L=13.3$ \AA\  and
a "zig-zag" (zz) structure at $L=14.8$ \AA.
Energetically, these (relaxed) structural isomers are found to
be local minima with higher energies (and slightly lower conductance) 
than the corresponding bc configurations (see Fig. 1a).
The zz isomer and bc structure
(as well as the lc) become approximately degenerate
at $L=15.8$ \AA, and upon further elongation
(i.e. for $L=16.3$ \AA) both the zz and the bc convert to 
a linear chain.
Note that although all these isomers
(rh, zz, bc, lc) are found to be local minima
for a free Au$_4$ cluster,
 the energetic ordering of structural isomers
 observed here for the four-atom NW 
 cannot be deduced from that 
 of the free cluster\cite{Au4}.

{\it lc$\rightarrow$breaking} ($L\geq 16.3$ \AA).\ \ \ 
With continued pulling 
 the lc wire shows
a tendency toward dimerization. At first
(for $16.3\leq L\leq 17.3$ \AA) the only
stable structures correspond to internal dimerization (id)
of the wire, with $d_{ii}<d_{ie}$\cite{nootti2}.
At the start of this elongation interval the conductance decreases only
slightly, with a marked drop at $L=17.3$ \AA. However, starting at
$L=17.3$ \AA\ an energetically competitive dimerization mode 
 of the lc wire emerges (being essentially
degenerate with the id structure at 17.3 \AA) 
where $d_{ii}>d_{ie}$\cite{nootti3},
with the conductance of this end-dimerized 
(ed) structure being $1g_0$.
 We note here
(see Fig. 1a) the markedly
higher conductance of the ed structure ($G=1g_0$) compared to
that of the corresponding id chain\cite{ourdimer}. Such end-dimerized
structures \cite{Oka99} are energetically favored for the rest of the
elongation process (that is for
$L>17.3$ \AA)\cite{nootti4}.
The force required for elongation during the 
end-dimerization stage is rather small
(i.e. $\sim$0.3 nN in the interval  17.6 \AA\ $\leq L\leq$ 18 \AA), 
and the conductance of the wire
decreases sharply  as eventual breaking is approached.

Throughout the structural evolution the electronic spectrum
in the wire region  is characterized by a dominant contribution of
atomic $d$ orbitals
 in the energy range $E_F-4.5\ eV\leq E\leq E_F-1 \ eV$,
with $s$, $p$, and $spd$ hybrids contributing  
at energies above and below this range. 
Examination of the local densities of states
 reveals that the main variations
in the spectrum
 in response to the mechanical elongation
occur predominantly in the interelectrode   spatial region
(the wire atoms, and the electrode atoms bound 
directly to them), involving mainly states in the 
vicinity of $E_F$.

To illustrate the nature of these states and their high 
sensitivity to structural variations we examine,
for selected atomic configurations developed during the
elongation process, orbital images (Fig. 2(b-d)) 
and their angular-momentum decompositions
about the wire atoms,
corresponding to electronic states  near $E_F$
(i.e. highest  occupied  molecular
orbitals, HOMO($j$), with the index
$j=-1,-2$ corresponding to the next-to-highest occupied orbitals
in descending order).
We observe that
for the ds structure (see Fig. 2b and the
configuration shown at the bottom left in Fig. 1a) the HOMO is of 
predominant $d_{xz}$ character ($p_x^{0.2}d_{xz}^{0.76}$, here and
in the following only the dominant $m$-components of $p$-and
$d$-characters are given) and
the HOMO(-1) is of strong $sp_z$ character ($s^{0.34}p_z^{0.65}d_{z^2}^{0.12}$)
on the strand atoms.
These states  evolve after the ds$\rightarrow$bc
transformation (see top left configuration in Fig. 1a) into  a
HOMO  with $s^{0.45}p_z^{0.50}$ ($p_x^{0.25}p_z^{0.25}d_{xz}^{0.32}$)
on  inner- (end-) atoms, and a HOMO(-1)
 with an almost pure (tilted) $d_{z^2}$ character on the
inner-atom  ($p_x^{0.03}d_{z^2}^{0.9}$)
and a ($s^{0.5}p_x^{0.08}d_{x^2-y^2}^{0.32}$) hybrid on the
end-atoms of the bent-chain (Fig. 2c).

The structural evolution during
the  bc$\rightarrow$lc stage and the formation
of dimerized configurations of the lc
are accompanied by reordering of the orbitals
near $E_F$ with the HOMO orbital on the
inner atoms of the wire acquiring a predominant
($p_x^{0.03}d_{xz}^{0.95}$) $d_{xz}$ character (see e.g.
Fig. 2d corresponding to the ed configuration);
 as evident the $d_{xz}$ components
on the two inner  atoms of the ed wire combine locally
in an antibonding manner, while in the (lower-energy) 
HOMO(-1) orbital
(not shown) they are in a local bonding orientation
with respect to each other.
The HOMO(-2) orbital (Fig. 2d) of the ed configuration
is characterized by a strong $s$ character on the inner-atoms
($s^{0.86}p_z^{0.11}d_{z^2}^{0.03}$)
combining to form a $\sigma$-like local bond,
 and an $sd$ hybridization
on the end-atoms of the chain ($s^{0.48}p_z^{0.06}d^{0.46}_{x^2-y^2}$).

The above hybridization patterns and in particular
the strong contribution from d-orbitals to directional
covalent-like bonding characteristics in the (interelectrode)
wire region, underlie and facilitate the structural
transformation mechanism described by us,
via stabilization of the sequence of strained
atomic configurations which evolve
in the elongation process. Consequently,
formation of such (several atom long) single-chain
nanostructures, reflected in an extended 
conductance plateau 
(with close to unit conductance)
before breaking\cite{Ya98} may not
be found for materials where the above
bonding patterns do not occur;
indeed extended single-chains were not
observed (theoretically\cite{nature}
and experimentally\cite{Ya98}) for a  
 simple  metal contact
(e.g. sodium).

Finally, we  remark that the investigations presented
here, which  deepen 
our insights into the structure, formation mechanism, and conductance
of atomic double-strand and single-strand 
nanocontacts between gold electrodes,
pertain to the type of tip-pulling
experiments reported  in the {\it first}
part of Ref. 3 (see Fig. 3 in Ref. 3), 
and they are {\it not} aimed at explaining the extraordinary
stability of monoatomic thick  wires bridging  two sections
of a suspended  thin gold film, where interatomic distances up
to 3.5 \AA\ -- 4 \AA\ were reported from HRTEM images 
(see the {\it second} part of Ref. 3, particularly Fig. 4 therein).
Indeed, the formation  of  wires in the latter experiments
 involves an initial
intense electron bombardment  to perforate holes in the film and
subsequent thinning of the formed "nanobridges" driven  by stress and strain
relaxation processes; additionally,
the nature of such experiments
 differs greatly from the controlled tip-pulling
experiments, on which we focus here.
We remark  that none of the 
theoretical studies reported to date
 \cite{tosatti,Oka99,zigzag,springborg} have been able to 
consistently
explain the  abovementioned 
anomalously long interatomic distances in
the nanobridges \cite{ugarte},
indicating that new theoretical considerations  are 
 needed, including examinations of possible effects caused
 by light impurity
atoms, which while affecting the structure of the
nanowire may not have enough contrast to be visible in 
HRTEM images.\cite{final}

This research is supported by the U.S. DOE
 and the Academy of Finland. 
Calculations were performed on an
IBM SP2  at the Georgia Tech
Center for Computational Materials Science,
and on a Cray T3E
 at the National Energy Research Scientific
Computing Center (NERSC) at Berkeley,
CA.

\vfil\eject
\section*{FIGURES}

FIG. 1. (a)
Energies ($\Delta E$, in unit of eV plotted at the bottom, relative to the
initial ds configuration at $L=11.5$ \AA) and conductances
($G$, in unit of $g_0$, plotted at the top) of Au NWs
versus the distance between the outermost layers of the opposing
electrodes, $L$; the double-strand (ds, squares), bent-chain
(bc, triangles), and linear-chain (lc, circles)
intervals are marked at the top. The values 
for the rhombic (rh) isomer at $L=13.3$ \AA\
and for the zig-zag (zz) isomers at $L=14.8$ \AA\
and 15.8 \AA\ are marked by crosses, and
those for the end-dimer configurations at $L=17.3$ \AA, 17.6 \AA\ and
18 \AA\ are depicted by stars. The elongation
force between consecutive structures may be estimated
by the corresponding slope $\Delta E/\Delta L$. 
The atomic configurations
 (with atom indices) shown as insets to (a) correspond to
the  ds (left, bottom), bc (left, top), and lc (right) structures at
$L=12.6$ \AA, 13.3  \AA\ and 16.3 \AA\ respectively. 
(b) atomic configurations for the rh isomer at $L=13.3$ \AA\ 
($d(1,2)=2.76$ \AA, $d(2,3)=2.59$ \AA, and $\angle(1,2,4)=124^\circ$),
and for the zz isomer at $L=14.8$ \AA\
($d(1,2)=d(3,4)=2.55$ \AA, $d(2,3)=2.52$ \AA\ and $\angle(1,2,3)=119^\circ$).

FIG. 2 (color)
(a) Iso-surface images of the magnitudes
of the first ($0.97g_0$, top)
and second ($0.7g_0$, bottom) transmitted 
eigenchannels for the highly-strained 
ds configuration (at $L=12.6$ \AA).
(b and c) HOMO (top) and HOMO(-1)
(bottom) orbitals corresponding to the 
highly-strained ds NW (in b)
and  to the bc configuration  at $L=13.3$ \AA\
(in c). (d) HOMO (top) and HOMO(-2) (bottom) orbitals
for the ed configuration at $L=17.3$ \AA.
In (a-d) Au atoms are depicted by
red spheres.
In (a) we display the entire atomic system
including parts of the jellium-slab
continuations of the
electrodes (used in the conductance 
calculations), and in (b-d) 
we focus on the vicinity of the 
nanowires. The viewing orientation
was chosen in each case to
enhance  visualization of the
orbital characters on the wire atoms.  


\begin{thebibliography}{99}
\bibitem{longds} For a relaxed longer ds wire
(i.e. two 4-atom-long strands with $d_{inter}=2.59$ \AA,
$d_{intra}=2.73$ \AA, and $d_{el-el}=11.73$ \AA),
we obtained $G=1.97 g_0$.
\bibitem{Au4}
The same LDA calculations
(Ref. 9)  for the free Au$_4$ cluster show that 
the ground state  is a planar
rhombus (rh) with $d(1,2)=2.70$ \AA\
and  $\angle(1,2,4)=121^\circ$.
 The zz isomer  with
$d(1,2)=d(3,4)=2.55$ \AA, 
$d(2,3)=2.60$ \AA\ and $\angle(1,2,3)=130^\circ$
(see atom numbering in Fig. 1b for rh and zz) is of higher
energy ($\Delta E=0.84$ eV), 
 and the  lc
 isomer, with
$d(1,2)=d(3,4)=2.52$ \AA\ and $d(2,3)=2.59$ \AA,
is of even higher energy
($\Delta E=1.34$ eV). A bc isomer
($d(1,2)=d(3,4)=2.51$ \AA, $d(2,3)=2.61$ \AA,
and $\angle(1,2,3)=126^\circ$)
is found to be the highest in energy
($\Delta E=1.40$ eV).
\end{thebibliography}
\end{document}